# Predictive Simulation of Interphases on Li Metal Surface

Xinyu Li[1,†], Jingxuan Ding[1,†], Daniel C. Hannah[1], Yumin Zhang[1], Qichao Hu[1], Kang Xu[1,*]

[1]SES AI, Woburn, MA, USA

† These authors contributed equally to this work.

* Corresponding author: kang.xu@ses.ai

## Abstract

Interphases remain the least understood components in advanced batteries. Although their properties dictate whether a new battery chemistry could perform as designed, there has never been a reliable way to predict what an interphase could arise from a new electrolyte system due to the absence of atomistic level knowledge about interphasial formation process. In this work, we attempt to develop a simulation method that can universally predict interphasial chemistries formed on Li metal surface, so that the electrolyte engineering would no longer need lengthy Edisonian approaches. By combining a transferable universal polarizable force field and a universal machine learning force field, we simulate interphasial chemistry across chemically diverse electrolyte formulations, and successfully replicate the experimental observation that fluorinated solvents promote the formation of LiF-rich interphases, whereas interphases of more organic origin arise from conventional carbonate-based electrolytes. By directly capturing these spontaneous interfacial reactions behind these interphasial chemistries, our simulations establish molecular-level relationships between electrolyte chemistry, salt concentration, decomposition pathways, and SEI properties, and opens a route toward universal and high-throughput predictive simulation of interphases that is the foundation for AI-driven electrolyte discoveries.

## Introduction

Lithium metal is the ultimate anode material that promises theoretical energy densities above 500 Wh $Kg^{-1}$ due to its high specific capacity (3860 $mAhg^{-1}$) and ultra-low electrochemical potential (-3.04 V vs standard hydrogen electrode). However, the practical use of lithium metal has been prevented by its extreme reactivity, which requires stabilization by an ideal interphase [1,2]. Known as the most important but least understood component in advanced batteries [3], interphases are formed via sacrificial decomposition of electrolyte components on electrode surfaces [4,5], which are ion-conducting but electron-insulating, and whose chemistry and formation process are still not fully understood despite decades of research. However, it has been well established that any efforts of developing electrolyte materials for new battery chemistries are essentially the design of interphases arising from these materials [6]. Thus far, our knowledge about interphasial chemistry on Li-metal has largely come from ex-situ analyses. For example, X-ray photoelectron spectroscopy (XPS) accompanied with etching techniques reveal that interphases often consist of chemically heterogeneous layers at nanometer-scale, and often buried by free electrolytes. Cryogenic electron microscopy further revealed previously inaccessible structural information, but they also highlight the sensitivity and complexity of native lithium-metal interfaces [7,8]. On the other hand, due to the absence of atomistic understanding, interphase engineering efforts have always been following a trial-and-error Edisonian approach, which is both time- and resource-consuming, and apparently far from being ready to support the emerging approaches of AI-driven materials discoveries that are hungry for high-throughput data [9]. While numerous AI models have been reported in the past few years [10-12], their capabilities have been mostly confined to the prediction of bulk properties such as ion conductivity, density and viscosity. This was because computationally, interphase formation is challenging because it involves coupled electron transfer, bond breaking and formation, ion transport, electrolyte solvation, and morphological evolution over length and time scales that exceed the capability of conventional ab initio molecular dynamics (MD). Classical force fields can be applied to larger length scales and longer timescales, but lack accuracy in describing spontaneous electrolyte decomposition and interphase growth. Developing a universal, transferable and scalable computational simulation method for accurate prediction of interphases constitutes a core foundation for the future AI4Materials.

Recent advances in machine learning force fields (MLFFs) provide a new route for atomistic simulation of reactive interfaces in detail, providing insights of the early-stage formation of SEI structures [13] and large-scale digital twins of SEI evolution [14]. It extends the capability of conventional ab initio molecular dynamics into unprecedented scales while maintaining quantum accuracy. Contrary to specific ML models, universal MLFFs trained on large quantum-mechanical datasets maintain quantum level accuracy yet greatly improve the chemical transferability to all involved species and systems [15-18]. For efficient predictions of SEI in various systems, this capability is particularly attractive because MLMD can therefore utilize the embedded chemical information of reactants (e.g. Li metal, salt and solvents) and explicit solid-liquid configurations in its training domain to accurately predict the formation of SEI structures.

In this work, we combine a transferable universal polarizable force field and a universal machine learning force field to simulate SEI formation across chemically diverse electrolyte formulations without developing specific reactive force fields for each electrolyte chemistry. Four representative electrolyte systems, LiFSI in FSA, LiFSI in EC/DMC/FEC, LiFSI in F5DEE, and LiFSI in SFL/TOL, were selected to compare how solvent chemistry and salt environment influence interfacial decomposition and LiF-rich SEI formation under a consistent simulation protocol. They were selected on the basis of their distinct chemical natures and functional group diversity: while EC/DMC/FEC stands for typical carbonate-systems mature in lithium-ion batteries (LIBs) based on graphite and silicon-carbon anodes, F5DEE and FSA represent ethereal and sulfonamide classes that were designed for Li metal surfaces especially. SFL/TOL, on the other hand, represents a new trend of eliminate carbonate solvents [19]. By analyzing interfacial structures, reaction counts, and LiF distributions, we are able to identify composition-dependent trends in SEI growth and provide mechanistic insight into electrolyte-controlled lithium-metal passivation. We successfully replicated the experimental observation that fluorinated solvents such as FSA and F5DEE promote the formation of LiF-rich SEI, whereas the conventional EC/DMC/FEC electrolyte produces a more organic-rich SEI. By directly capturing spontaneous interfacial reactions, our simulations establish molecular-level relationships between electrolyte chemistry, salt concentration, decomposition pathways, and SEI properties. This approach paves the way toward a scalable high-throughput predictive simulation of interphases on all electrode surfaces.

## Methods

### Compositions

Six electrolyte formulations were considered. Four baseline formulations were LiFSI in FSA, LiFSI in EC/DMC/FEC, LiFSI in F5DEE, and LiFSI in SFL/TOL, where LiFSI is lithium bis(fluorosulfonyl)imide, FSA is *N,N*-dimethylsulfamoyl fluoride, EC is ethylene carbonate, DMC is dimethyl carbonate, FEC is fluoroethylene carbonate, F5DEE is 2-[2-(2,2-difluoroethoxy)ethoxy]-1,1,1-trifluoroethane, SFL is sulfolane, and TOL is toluene. Two additional LiFSI in FSA compositions, with salt-to-solvent molar ratios of 1:6 and 1:10, were constructed to probe the effect of salt concentration. The final cell compositions are listed in Table. 1.

**Table 1. Bulk electrolyte compositions.**

| Electrolyte | Molecular ratio | Molecule counts | Electrolyte atoms | Total atoms |
|---|---|---|---|---|
| LiFSI/EC/DMC/FEC | 1:6.87:5.44:1.15 | 32 LiFSI, 220 EC, 174 DMC, 37 FEC | 4,978 | 7,678 |
| LiFSI/F5DEE | 1:6.1 | 34 LiFSI, 207 F5DEE | 4,894 | 7,594 |
| LiFSI/SFL/TOL | 1:3.5:3.12 | 48 LiFSI, 168 SFL, 150 TOL | 5,250 | 7,950 |
| LiFSI/FSA | 1:3.5 | 83 LiFSI, 291 FSA | 4,613 | 7,313 |
| LiFSI/FSA | 1:6 | 52 LiFSI, 312 FSA | 4,576 | 7,276 |
| LiFSI/FSA | 1:10 | 33 LiFSI, 330 FSA | 4,620 | 7,320 |

**Electrolyte construction and equilibration**

The bulk electrolyte structures were optimized with ByteFF-Pol, a graph neural network parameterized polarizable force field trained on quantum mechanical data [12]. Initial configurations were packed with the insert-molecules utility in GROMACS 2025.03 [20] into orthorhombic cells of dimensions 3.456 x 3.456 x 5.000 nm, where the x and y dimensions matched an optimized Li slab from UMA, a universal machine-learning interatomic potential [16], and the z dimension was our target thickness. Periodic boundary conditions were applied in all three dimensions, and each system was charge neutral.

Bulk equilibration was performed with OpenMM 8.4 [21] using a constrained anisotropic isothermal-isobaric ensemble (NPT). The lateral cell vectors were fixed at 3.456 nm, whereas the cell length along z was allowed to fluctuate under a Monte Carlo anisotropic barostat at 1 atm, 298 K. Before equilibration, each packed configuration was minimized for at most 1,000 iterations using an energy-gradient tolerance of 10 kJ $mol^{-1}$ $nm^{-1}$. The equations of motion were propagated with the OpenMM BAOAB-RESPA multiple-time-step Langevin integrator [22, 23] with a 2 fs timestep. Bonded and other rapidly varying forces were evaluated twice per outer step, whereas nonbonded, custom nonbonded, and multipolar forces were evaluated once per outer step. The Langevin collision frequency was 0.1 $ps^{-1}$. Each system was equilibrated for 5,000,000 steps (10 ns), and temperature, energy, stress tensor and coordinates were recorded every 5,000 steps. Density and cell-length statistics were calculated over the final 2 ns of each trajectory.

The mean bulk properties used in model construction are summarized in Table 2. The uncertainties are the standard deviations of the sampled values over the final 2 ns.

**Table 2. Bulk-electrolyte properties from ByteFF-Pol equilibration.**

| **Electrolyte** | **Mean density (g $mL^{-1}$)** | **Mean $L_z$ (nm)** |
|---|---|---|
| LiFSI/EC/DMC/FEC | 1.254 +/- 0.009 | 4.984 |
| LiFSI/F5DEE | 1.369 +/- 0.010 | 5.021 |
| LiFSI/SFL/TOL | 1.196 +/- 0.007 | 4.994 |
| LiFSI/FSA (1:3.5) | 1.456 +/- 0.008 | 5.015 |
| LiFSI/FSA (1:6) | 1.384 +/- 0.008 | 4.961 |
| LiFSI/FSA (1:10) | 1.336 +/- 0.008 | 5.008 |

**Lithium metal slab and interface construction**

The lithium electrode was constructed from body-centered-cubic Li with the [001] direction normal to the interface. A periodic 10 x 10 in-plane supercell was relaxed with the UMA-s-1p2 machine-learning interatomic potential, using the OC25 task in FAIR-Chem 2.21.0 [24]. The atomic positions and cell were relaxed with the fast inertial relaxation engine (FIRE) [25] and an ASE Frechet cell filter [26] until the maximum force was below 0.001 eV $Å^{-1}$. The optimized conventional cell lattice constant was 3.456 Å.

The Li slab contained 27 atomic layers and a total of 2,700 atoms. The Li slab was placed at the bottom of the cell from 0 to 45.4 Å. The bottom two layers of Li (200 atoms, 0.5 <= z <= 2.3 Å) were fixed during interfacial dynamics and all other Li atoms were mobile. We used this fixed configuration to eliminate the surface effects of Li under vacuum due to the periodic boundary condition.

The initial and final structures of our SEI simulations are shown in Figure 1. For each electrolyte, the final bulk configuration was first wrapped in the lateral directions. Molecules intersected by the periodic boundary normal to the interface were then unfolded as intact molecules along z, thereby avoiding artificial bond discontinuities when the periodic liquid was converted into a slab geometry. The electrolyte was translated so that its minimum z coordinate was 2.0 A above the uppermost Li layer.

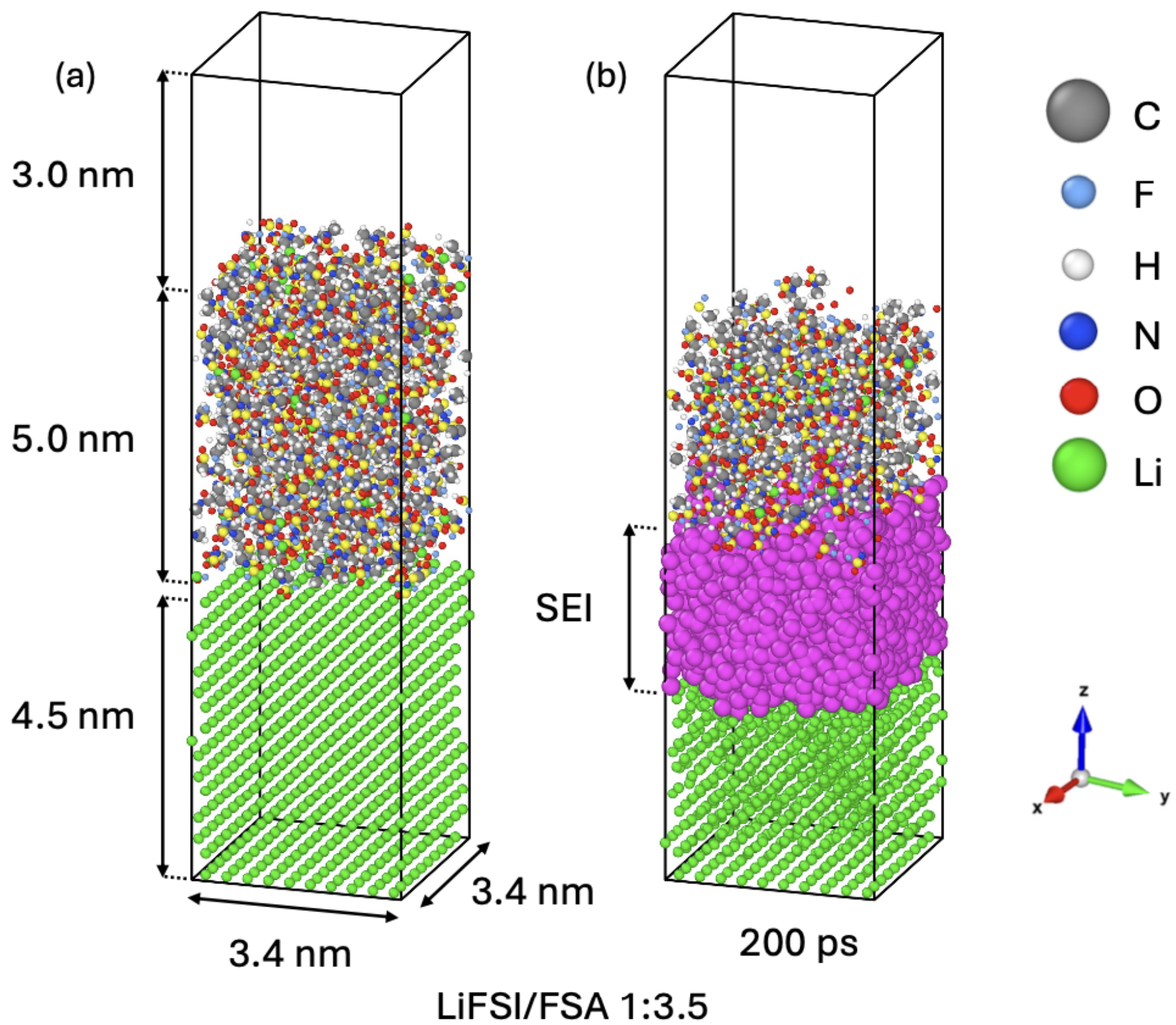

**Figure 1. Construction and evolution of the SEI.** Representative LiFSI/FSA 1:3.5 simulation cell before and after interfacial reaction. The initial model contains a Li metal slab, an approximately 5.0 nm equilibrated electrolyte region, and a 3.0 nm vacuum layer in a periodic cell with fixed lateral dimensions of approximately 3.4 nm × 3.4 nm. The Li slab and electrolyte region has a 2 Å gap. After 200 ps, a LiF-rich solid-electrolyte interphase forms at the Li/electrolyte interface. Atomic colors denote C, F, H, N, O, and Li as indicated.

The periodic cell was subsequently extended in z to introduce a 30 Å vacuum region above the electrolyte. A one-sided harmonic wall was placed 1.5 Å above the largest initial z coordinate of any electrolyte atom. The resulting cell dimensions and wall positions are given in Table 3. Periodic boundary conditions remained active in all three dimensions. The vacuum therefore separated the electrolyte from the periodic image of the lower Li surface. No additional geometry optimization was performed after assembling the Li/electrolyte/vacuum cells.

**Table 3. Interfacial cell dimensions and upper-wall positions.**

| **Electrolyte** | **Cell dimensions ($Å^3$)** | **Harmonic-wall position $z_w$ (Å)** |
|---|---|---|
| LiFSI/EC/DMC/FEC | 34.565 x 34.565 x 132.711 | 102.711 |
| LiFSI/F5DEE | 34.565 x 34.565 x 136.483 | 106.483 |
| LiFSI/SFL/TOL | 34.565 x 34.565 x 133.241 | 103.241 |
| LiFSI/FSA (1:3.5) | 34.565 x 34.565 x 132.903 | 102.903 |
| LiFSI/FSA (1:6) | 34.565 x 34.565 x 133.079 | 103.079 |
| LiFSI/FSA (1:10) | 34.565 x 34.565 x 132.669 | 102.669 |

The harmonic wall acted independently on every non-slab atom, including $Li^+$ ions. For an affected atom *i*, its contribution to external potential and force were

$$U_i(z_i) = \frac{1}{2} k \left[\max(0, z_i - z_w)\right]^2$$

$$F_{i,z} = -k \cdot \max(0, z_i - z_w)$$

with $k$=1.0 eV Å$^{-2}$. The corresponding force was zero for $z_i <= z_w$ and $F_{i,z} = -k(z_i - z_w)$ for $z_i > z_w$. Thus, the wall opposed penetration into the vacuum but exerted no force within the nominal liquid region. This boundary treatment was adapted from Takenaka *et al.* [13].

### Molecular dynamics

Interfacial molecular dynamics was performed with UMA-s-1p2 using the OC25 task in FAIR-Chem 2.21.0. OC25 was selected because it includes explicit solid-liquid configurations, solvents, and ions in its training domain. Energies and forces were interfaced to the Atomic Simulation Environment (ASE 3.29.0) using PyTorch 2.8.0 with CUDA 12.8 support. The calculations used fixed atom numbers and fixed cell vectors, with a local Langevin heat bath (NVT) at a target temperature of 298 K. Because the stochastic thermostat acted only within a spatially selected region, this protocol should be understood as locally thermostatted fixed-cell dynamics rather than an exact global canonical sampler. The integration time step was 1.0 fs, and each production trajectory was 200 ps (200,000 steps).

Temperature control used the quasi-symplectic Langevin propagator [27] implemented in ASE. A friction coefficient of 0.1 ps$^{-1}$ was applied only to non-fixed atoms whose coordinates in the initial reference structure satisfied $0 <= z <= 70$ Å. This region comprised the mobile Li slab and the electrolyte adjacent to the metal surface. Atoms outside this region evolved without direct stochastic or frictional forces. This avoids directly thermostating the electrolyte/vacuum boundary while allowing energy exchange through interatomic interactions with the thermostatted region. Center-of-mass correction was disabled during Langevin propagation. Initial velocities were drawn from a Maxwell-Boltzmann distribution at 298 K with exact initial kinetic-temperature rescaling. The net linear momentum was then removed, after which the momenta of the fixed Li atoms were set to zero. Continuation runs retained the coordinates and momenta stored in the final frame of the preceding trajectory segment rather than reinitializing velocities.

The harmonic-wall contribution was added to total energy and atomic forces but not to virial stress. Stress was not used for thermodynamic interpretation. UMA inference was distributed across two GPUs for production calculations. The use of multiple GPUs changes only the force-evaluation throughput and does not alter the simulated Hamiltonian or integration parameters.

## Simulation results and discussion

### Electrolyte-dependent SEI morphology and thickness

The SEI structures obtained after 200 ps of interfacial molecular dynamics simulations are shown in Figure 2. The six electrolyte formulations generate markedly different interphase morphologies, demonstrating a strong dependence of SEI formation on electrolyte chemistry and salt concentration. Among the simulated systems, the LiFSI/FSA electrolytes form substantially thicker SEI than the other formulations. As summarized in Table 4, the SEI thicknesses for LiFSI/FSA with salt-to-solvent molar ratios of 1:3.5, 1:6, and 1:10 are 30.85, 28.33, and 23.39 Å, respectively. In comparison, substantially thinner SEIs are obtained for LiFSI/F5DEE (9.77 Å), LiFSI/SFL/TOL (9.35 Å), and LiFSI/EC/DMC/FEC (7.63 Å).

Within the LiFSI/FSA electrolyte series, increasing salt concentration leads to a systematic increase in SEI thickness. The thickness increases from 23.39 Å at a LiFSI:FSA ratio of 1:10 to 28.33 Å at 1:6 and 30.85 Å at 1:3.5. Interestingly, the LiF amount itself varies only weakly across these concentrations. The mole fraction of LiF remains approximately 26–27%, while the number density of LiF varies only from 0.00451 to 0.00473 $Å^{-3}$. Thus, the increased SEI thickness at higher salt concentration shows little correlation with the local density of LiF. Instead, increasing the concentration of LiFSI leads to greater participation of FSI anions during SEI formation. This results in a higher fraction of salt-derived inorganic products in the SEI.

These observations illustrate an important advantage of reactive MLFF-MD simulations: interphase growth is captured simultaneously with underlying reactions. The simulations directly describe bond breaking and formation, ionic transport, and structural evolution at the heterogeneous Li/electrolyte interface. As a result, differences in SEI morphology and thickness can be connected to reactions at the molecular scale.

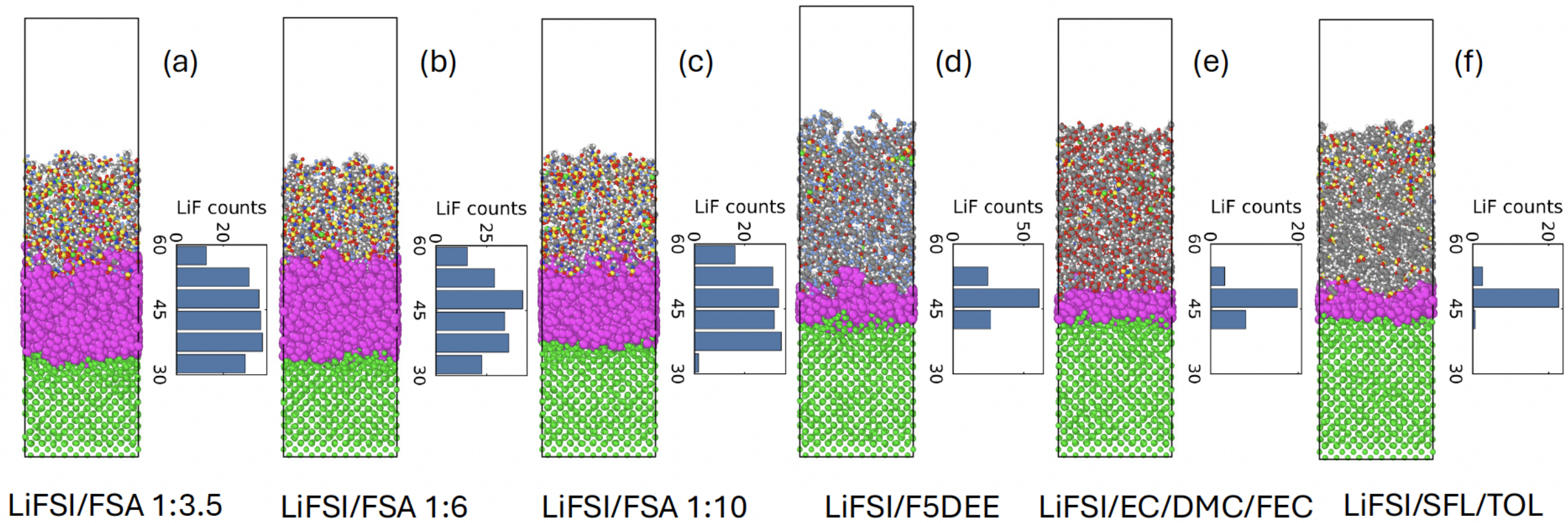

**Figure 2. Composition-dependent LiF formation at Li/electrolyte interfaces.** Final configurations from molecular dynamics at 200 ps for six electrolyte formulations: (a) LiFSI/FSA 1:3.5, (b) LiFSI/FSA 1:6, (c) LiFSI/FSA 1:10, (d) LiFSI/F5DEE, (e) LiFSI/EC/DMC/FEC, and (f) LiFSI/SFL/TOL. The adjacent horizontal histograms show the distribution of LiF counts along the interfacial direction, binned between 30 and 60 Å. The Li metal slab is shown in green, SEI in magenta, and electrolyte atoms are colored by element.

**Table 4. Summary of SEI thickness and compositions**

| Electrolyte | Molecular ratio of reacted molecules | SEI thickness (Å) | LiF density in SEI (Å$^{-3}$) | Mole fraction of LiF (%) | Mole fraction of organic components (%) |
|---|---|---|---|---|---|
| LiFSI/EC/DMC/FEC | FSI:EC:DMC:FEC = 1:1.50:0.58:0.58 | 7.63 | 0.00318 | 22.86 | 34.56 |
| LiFSI/F5DEE | FSI:F5DEE = 1:0.84 | 9.77 | 0.00599 | 32.61 | 11.03 |
| LiFSI/SFL/TOL | FSI:SFL:TOL = 1:1.15:0 | 9.35 | 0.00233 | 29.79 | 5.68 |
| LiFSI/FSA | FSI : FSA = 1:3.70 | 30.85 | 0.00451 | 26.61 | 18.41 |
| LiFSI/FSA | FSI : FSA = 1:4.89 | 28.33 | 0.00467 | 26.35 | 18.10 |
| LiFSI/FSA | FSI : FSA = 1:10.42 | 23.39 | 0.00473 | 26.75 | 22.26 |

**Composition of simulated SEI**

Figure 3 compares the SEI compositions formed from the six electrolyte formulations and presents representative structures of major SEI components. The simulations reveal substantial differences in the ratio between inorganic and organic components depending on electrolyte chemistry and salt concentration.

A characteristic feature of the LiFSI/FSA systems is the formation of $Li_xSO_2NC_2H_6$ and $Li_xNC_2H_6$. Their molar ratio is approximately 1:2, suggesting that approximately one-third of the reacted FSA molecules serve solely as fluorine donors without further decomposition. Our simulation is consistent with previous experimental observation that FSA is a strong fluorine donor [29]. In addition, the C-N bond in $Li_xNC_2H_6$ is sufficiently stable that no C-N bond breaking is observed within the 200 ps simulation timescale. The stability of $Li_xNC_2H_6$ prevents the formation of organic carbonates, generating a more inorganic SEI.

The conventional LiFSI/EC/DMC/FEC electrolyte exhibits a distinctly different SEI composition. Among the six formulations, it produces the lowest LiF mole fraction and the highest organic-component mole fraction. Its LiF number density, 0.00318 Å$^{-3}$, is also lower than those of the LiFSI/FSA and LiFSI/F5DEE systems. These results indicate that decomposition of the carbonate solvents contributes substantially to SEI formation and produces an organic-rich interphase. In contrast, LiFSI/F5DEE produces the highest LiF mole fraction, and the highest LiF number density, 0.00599 Å$^{-3}$. It also exhibits a low organic-component mole fraction of 11.03%. These characteristics are associated with the high fluorine content of F5DEE and its extensive defluorination during interfacial reduction. These computational observations agree well with experimental characterizations of F5DEE electrolytes [30-32].

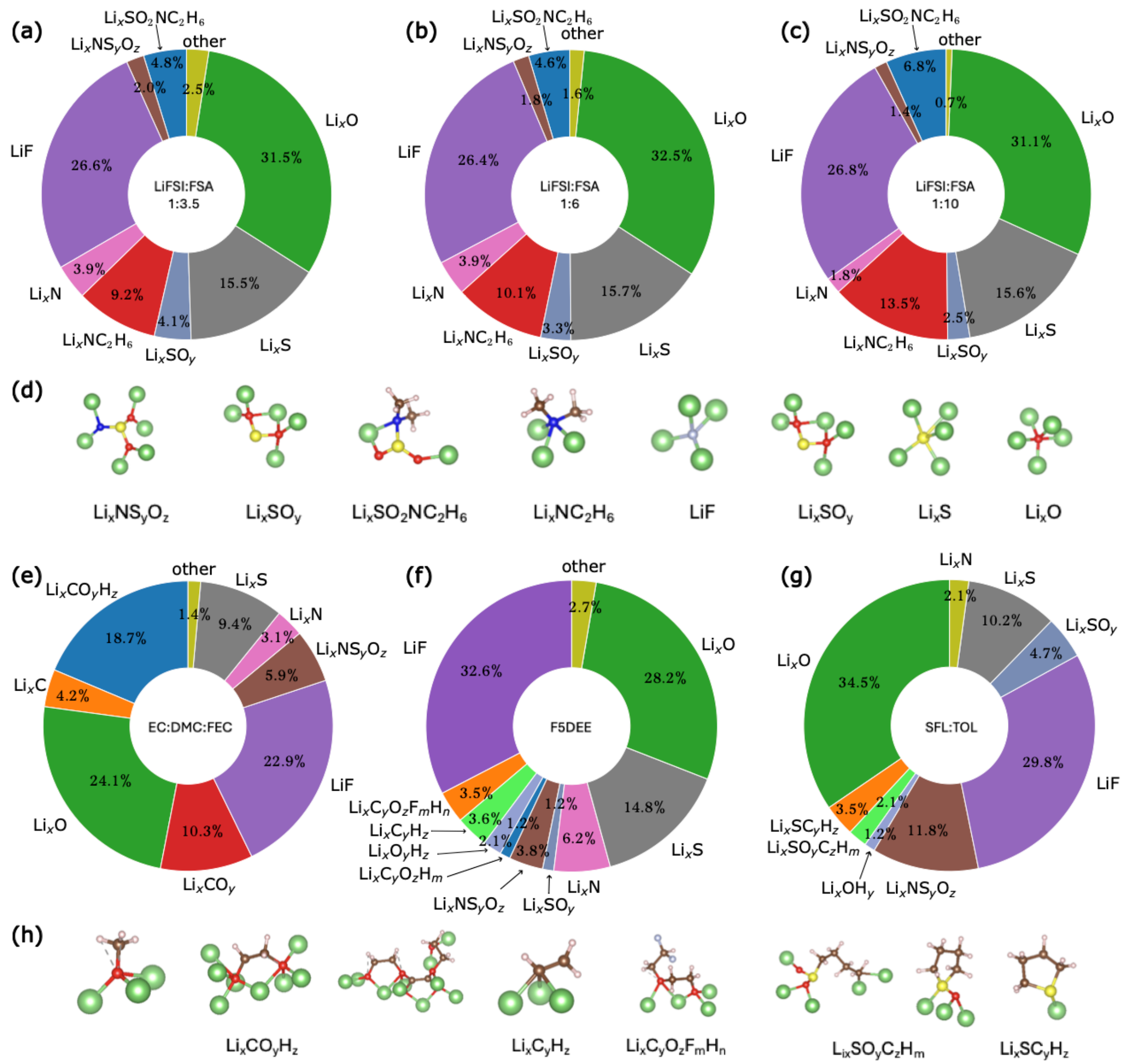

**Figure 3. Compositions and component structures of SEI.** Mole fractions of SEI components at 200 ps for six electrolyte formulations: (a) LiFSI/FSA (1:3.5), (b) LiFSI/FSA (1:6), (c) LiFSI/FSA (1:10), (e) LiFSI/EC/DMC/FEC, (f) LiFSI/F5DEE, and (g) LiFSI/SFL/TOL. Representative atomic structures of the major SEI components are shown in (d) and (h).

The LiFSI/SFL/TOL system forms a particularly thin SEI while maintaining a relatively high LiF mole fraction. Its SEI thickness is 9.35 Å, while LiF accounts for 29.79% of SEI components. Furthermore, the organic-component mole fraction is only 5.68%, the lowest among the six formulations. These results indicate that LiFSI/SFL/TOL produces a thin, relatively inorganic-rich interphase within the simulated 200 ps timescale. This increased LiF presence in interphases could directly result from the effect of diluent molecules, which forces F-donating molecules and anions into the solvation sheath of Li+ and form a tighter solvation cluster for reactions at anode.

**Molecular reaction pathways during SEI formation**

The reactive nature of the MLFF-MD simulations enables the molecular origins of these differences in SEI composition to be directly examined. Representative reaction pathways for FSA, EC, DMC, FEC, F5DEE, and SFL are summarized in Figure 4.

The LiFSI/F5DEE system provides an example of extensive solvent-derived LiF formation. Reaction pathway analysis shows that all five fluorine atoms in an F5DEE molecule can ultimately be released and contribute to LiF formation at the lithium metal surface. The C–F bonds at the –$CF_2$ end undergo reduction first, followed by those at the –$CF_3$ end. This molecular-level pathway explains the high LiF mole fraction observed in the SEI of LiFSI/F5DEE electrolyte.

The simulations also reveal a distinct role for TOL in the LiFSI/SFL/TOL electrolyte. No TOL reduction is observed during our MD simulations, indicating that TOL remains comparatively stable at the lithium metal interface within the simulated timescale. TOL therefore behaves primarily as a diluent. The thin SEI, relatively high LiF fraction, and low fraction of organic decomposition products suggest LiFSI/SFL/TOL as a promising candidate for lithium metal batteries.

These chemical reactions emerge directly from the atomic interactions during the MD simulations rather than being imposed through predefined reaction templates. This approach therefore does not require prior knowledge of decomposition pathways, which is particularly advantageous for SEI formation where multiple reactions can occur simultaneously in complex interfacial environments.

**Competitive salt and solvent reduction and interfacial passivation**

Beyond identifying individual reaction pathways, MLFF-MD enables statistical comparison of salt- and solvent-derived reactions and links them to SEI composition. Figure 5 tracks the numbers of reacted FSI anions and solvent molecules as a function of simulation time.

The LiFSI/FSA concentration series provides insight into the effect of salt concentration on competitive interfacial reduction. As shown in Figure 5 (a), (b) and (c), the number of reacted FSA solvent molecules remains nearly constant around 120, indicating that the reduction of FSA is insensitive to the salt concentration. In contrast, the number of reacted FSI anions is proportional to salt concentration. At all three concentrations, FSA dominates the reduction reactions at the lithium metal anode surface. The increased participation of FSI at higher salt concentrations provides an explanation for the increasingly inorganic-rich SEI observed in the LiFSI/FSA simulations.

According to Figure 5 (d), EC is the primary solvent that decomposes during the formation of initial SEI, resulting in a higher fraction of organic components in the SEI. In LiFSI/F5DEE electrolyte, FSI is slightly more prone to reduction than F5DEE, indicating substantial anion participation in the formation of the highly fluorinated interphase. For LiFSI/SFL/TOL, the molar ratio of reacted SFL to FSI is approximately 1:1, indicating that SEI formation in the LiFSI/SFL/TOL electrolyte lies near the transition between solvent-dominated and salt-dominated mechanisms.

**Figure 4. Reaction mechanisms of electrolyte solvents in the formation of SEI.** The reaction mechanisms of six solvents at the interface between lithium metal anode and electrolyte: (a) FSA, (b) EC, (c) DMC, (d) FEC, (e) F5DEE and (f) SFL. Due to the existence of various coordinate environments inside SEI, each solvent can undergo multiple reaction pathways. Only the predominant reaction pathway for each solvent molecule is plotted here.

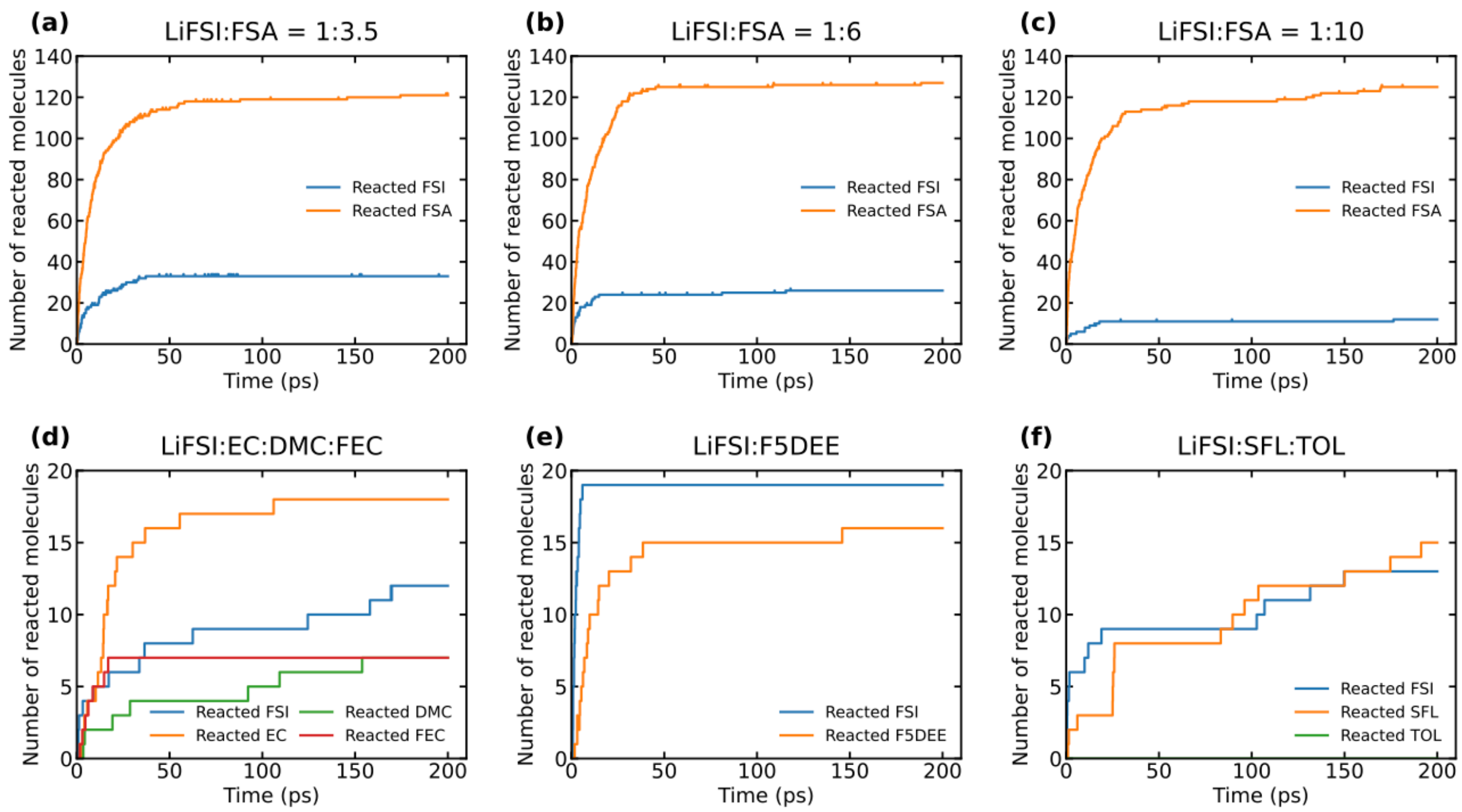

**Figure 5. Time evolution of the number of reacted anions and solvents.** The competition between FSI and solvents are studied in six electrolyte formulations: (a) LiFSI/FSA 1:3.5, (b) LiFSI/FSA 1:6, (c) LiFSI/FSA 1:10, (d) LiFSI/EC/DMC/FEC, (e) LiFSI/F5DEE and (f) LiFSI/SFL/TOL. The rapid increase in the number of reacted molecules, followed by a plateau, supports the proposed passivation mechanism.

These results demonstrate that the resulting SEI is governed not simply by the initial electrolyte composition, but by the competitive reduction kinetics of individual electrolyte components at the lithium metal surface.

## Conclusion

In this work we showed that scalable predictive simulation of interphases can be realized computationally by combining polarizable and machine-learning force fields. The further refining of this approach could lead to a universal tool that not only supports the AI-driven materials discovery for battery and electrolyte, but in broader context any processes that involve interfacial or interphasial chemistries, such as catalysis and electrosynthesis.

## References

[1] Lin, Dingchang, Yayuan Liu, and Yi Cui. "Reviving the lithium metal anode for high-energy batteries." Nature nanotechnology 12.3 (2017): 194-206.

[2] Cheng, Xin-Bing, et al. "Toward safe lithium metal anode in rechargeable batteries: a review." Chemical reviews 117.15 (2017): 10403-10473.

[3] Winter, Martin. "The solid electrolyte interphase–the most important and the least understood solid electrolyte in rechargeable Li batteries." *Zeitschrift für physikalische Chemie* 223.10-11 (2009): 1395-1406.

[4] Xu, Kang. "Nonaqueous liquid electrolytes for lithium-based rechargeable batteries." Chemical Reviews-Columbus 104.10 (2004): 4303-4418.

[5] Xu, Kang. "Electrolytes and interphases in Li-ion batteries and beyond." Chemical reviews 114.23 (2014): 11503-11618.

[6] Xu, Kang. *Electrolytes, Interfaces and Interphases: Fundamentals and Applications in Batteries*. Royal Society of chemistry, 2023.

[7] Li, Yuzhang, et al. "Atomic structure of sensitive battery materials and interfaces revealed by cryo–electron microscopy." Science 358.6362 (2017): 506-510.

[8] Zhang, Zewen, et al. "Capturing the swelling of solid-electrolyte interphase in lithium metal batteries." Science 375.6576 (2022): 66-70.

[9] Hannah, Daniel, et al. "Searching for ideal electrolytes in the molecular universe." *The Electrochemical Society Interface* 34.2 (2025): 35-38.

[10] Gong, Sheng, et al. "A predictive machine learning force-field framework for liquid electrolyte development." *Nature Machine Intelligence* 7.4 (2025): 543-552.

[11] Wang, Feng, et al. "Domain oriented universal machine learning potential enables fast exploration of chemical space of battery electrolytes." Nature Communications 17.1 (2025): 1226.

[12] Zheng, Tianze, et al. "Bridging quantum mechanics to liquid properties via a universal organic force field." *Nature Communications* (2026).

[13] Takenaka, Norio, et al. "Machine learning force field molecular dynamics simulation of SEI formation on lithium metal." npj Computational Materials (2026).

[14] Ding, Jingxuan, et al. "Coupled reaction and diffusion governing interface evolution in solid-state batteries." arXiv preprint arXiv:2506.10944 (2025).

[15] Chen, Chi, and Shyue Ping Ong. "A universal graph deep learning interatomic potential for the periodic table." Nature Computational Science 2.11 (2022): 718-728.

[16] Wood, Brandon, et al. "UMA: A family of universal models for atoms." Advances in Neural Information Processing Systems 38 (2026): 129391-129427.

[17] Zeni, Claudio, et al. "A generative model for inorganic materials design." Nature 639.8055 (2025): 624-632.

[18] Deng, Bowen, et al. "CHGNet as a pretrained universal neural network potential for charge-informed atomistic modelling." Nature Machine Intelligence 5.9 (2023): 1031-1041.

[19] Dahn, Jeff R. "Understanding and Eliminating Capacity Loss in LFP/Graphite and LFP/Si Cells Operated at High Temperature." *23rd International Meeting on Lithium Batteries*, 15 June 2026, Palais des congrès de Montréal, Montréal, Québec, Canada. Conference presentation.

[20] Abraham, Mark James, et al. "GROMACS: High performance molecular simulations through multi-level parallelism from laptops to supercomputers." *SoftwareX* 1 (2015): 19-25.

[21] Eastman, Peter, et al. "OpenMM 8: molecular dynamics simulation with machine learning potentials." *The Journal of Physical Chemistry B* 128.1 (2024): 109-116.

[22] Tuckerman, M. B. B. J. M., Bruce J. Berne, and Glenn J. Martyna. "Reversible multiple time scale molecular dynamics." *The Journal of chemical physics* 97.3 (1992): 1990-2001.

[23] Lagardere, Louis, Félix Aviat, and Jean-Philip Piquemal. "Pushing the limits of multiple-time-step strategies for polarizable point dipole molecular dynamics." *The journal of physical chemistry letters* 10.10 (2019): 2593-2599.

[24] Sahoo, Sushree Jagriti, et al. "The Open Catalyst 2025 (OC25) dataset and models for solid-liquid interfaces." AI4X {\textendash} Accelerate Conference 2026. 2025.

[25] Bitzek, Erik, et al. "Structural relaxation made simple." Physical review letters 97.17 (2006): 170201.

[26] Hjorth Larsen, Ask, et al. "The atomic simulation environment—a Python library for working with atoms." Journal of Physics: Condensed Matter 29.27 (2017): 273002.

[27] Vanden-Eijnden, Eric, and Giovanni Ciccotti. "Second-order integrators for Langevin equations with holonomic constraints." Chemical physics letters 429.1-3 (2006): 310-316.

[28] Weijiang, Zhe, Mingjun, et al. FSI-inspired solvent and "full fluorosulfonyl" electrolyte for 4 V class lithium-metal batteries. Energy Environ. Sci. 2020; 13 (1): 212–220.

[29] Yu, Z., Rudnicki, P.E., Zhang, Z. et al. Rational solvent molecule tuning for high-performance lithium metal battery electrolytes. Nat Energy 7, 94–106 (2022).

[30] Sha, Dacheng, et al. Evolution and Interplay of Lithium Metal Interphase Components Revealed by Experimental and Theoretical Studies. Journal American Chemical Society 1 May 2024; 146 (17): 11711–11718.

[31] Zehao Cui, Zhiao Yu, Hao Lyu, Zhenan Bao, Arumugam Manthiram; Resolving Electrolyte Decomposition Products in Gas, Liquid, and Solid Phases in Lithium–Metal Batteries. ACS Energy Lett. 8 August 2025; 10 (8): 3827–3833.

[32] Tan, Sha, et al. "Synchronized breathing in anion-derived interphases." *ACS Energy Letters* 10.8 (2025): 3746-3754.